\documentclass[a4paper,
               keeplastbox,   % flushend option: not to un-indent last line in References
               ]{jacow}
\usepackage{pdfpages,multirow,ragged2e} %
\makeatletter%
	\ifboolexpr{bool{xetex}}
	 {\renewcommand{\Gin@extensions}{.pdf,%
	                    .png,.jpg,.bmp,.pict,.tif,.psd,.mac,.sga,.tga,.gif,%
	                    .eps,.ps,%
	                    }}{}
\makeatother

\ifboolexpr{bool{xetex} or bool{luatex}} % test for XeTeX/LuaTeX
 {}                                      % input encoding is utf8 by default
 {\usepackage[utf8]{inputenc}}           % switch to utf8

\usepackage[english]{babel}

\ifboolexpr{bool{jacowbiblatex}}%
 {%
  \addbibresource{jacow-test.bib}
  \addbibresource{biblatex-examples.bib}
 }{}
\begin{document}

\title{Injection Feedback for a Storage Ring}%\NoCaseChange{JACoW}

\author{A. Moutardier, N. Delerue, C. Bruni, I. Chaikovska, S. Chancé, E.E. Ergenlik,\\ V. Kubytskyi, H. Monard, Université Paris-Saclay, CNRS/IN2P3, IJCLab, Orsay}
	
\maketitle

\begin{abstract}

We report on an injection feedback scheme for the ThomX storage ring project. ThomX is a 50-MeV-electron accelerator prototype which will use Compton backscattering in a storage ring to generate a high flux of hard X-rays. Given the slow beam damping (in the ring), the injection must be performed with high accuracy to avoid large betatron oscillations. 
A homemade analytic code is used to compute the corrections that need to be applied before the beam injection to achieve a beam position accuracy of a few hundred micrometers in the first beam position monitors (BPMs). In order to do so the code needs the information provided by the ring's diagnostic devices.
The iterative feedback system has been tested using MadX simulations. Our simulations show that a performance that matches the BPMs' accuracy can be achieved in less than 50 iterations in all cases. Details of this feedback algorithm, its efficiency and the simulations are discussed.
\end{abstract}

\section{ThomX}

ThomX is a 50-MeV-electron accelerator using Compton backscattering to generate a high X-ray flux.

For general information about ThomX see paper~\cite{loss_map} and the project's TDR~\cite{TDR}. In this paper we will focus on the end of the transfer line (TL), used to adapt the beam for the ring, and the ring itself. The lattice used in MadX is slightly different from the one used in the TDR.

\subsection{Injection}

\begin{figure*}[!bth]
    \centering
    \includegraphics*[width=0.9\textwidth]{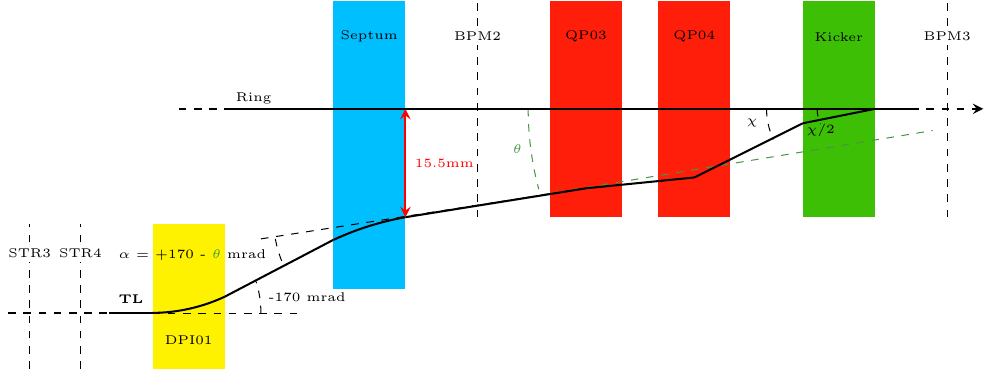}
    \caption{Sketch of the injection of ThomX. STR mean steerer, DPI mean injection dipole, BPM mean Beam Position Monitor, QP mean quadrupole.}
    \label{fig:schema_injection}
\end{figure*}

Once the beam is ready to be injected in the ring, an injection dipole propagates it through a septum to the ring.
A fast kicker in the ring is used to correct the injection angle and allow the beam to propagate following the optimal ring orbit (see Fig. \ref{fig:schema_injection}).

According to simulations, the \textbf{$\theta$} angle at the exit of the septum is \textbf{$\theta$}~=~\SI{11}{\milli\radian} for a kicker's kick of $\chi$~=~\SI{-13}{\milli\radian}.

\section{Injection simulation}

The injection in the ThomX's ring is simulated using MadX \cite{MadX}.

MadX is a code developed by the CERN to compute, among other things, the tracking of particles in accelerators.

On MadX a line %(called TL\_1\_turn)
has been created to simulate the propagation in the TL and in the first ring turn.

To simulate the off-axes propagation, in both quadrupoles between the septum and the kicker, during the injection a change of frame is done at the exit of the septum.
The reference frame of the particles at this localisation change from the beam reference frame to the reference frame of the ring.
With the right choice of values, the beam reference frame and the ring one will overlap in the kicker and stay the same further away in the ring.

The change of frame used in the code is :
\begin{itemize}
    \item X $\rightarrow$ X - \SI{15.5}{\milli\meter}
    \item PX $\rightarrow$ PX + \textbf{$\theta$}
\end{itemize}

Moreover, the kicker is simulated by 2 zero-length kickers at the beginning and the end of the real one with a kick of $\chi/2$ each.

\section{Injection feedback code}
\subsection{Principle}

The goal is to measure some beam parameters in the ring and to compute the corrections needed to improve the injection.

\begin{figure}[!htb]
   \centering
   \includegraphics*[width=0.75\columnwidth]{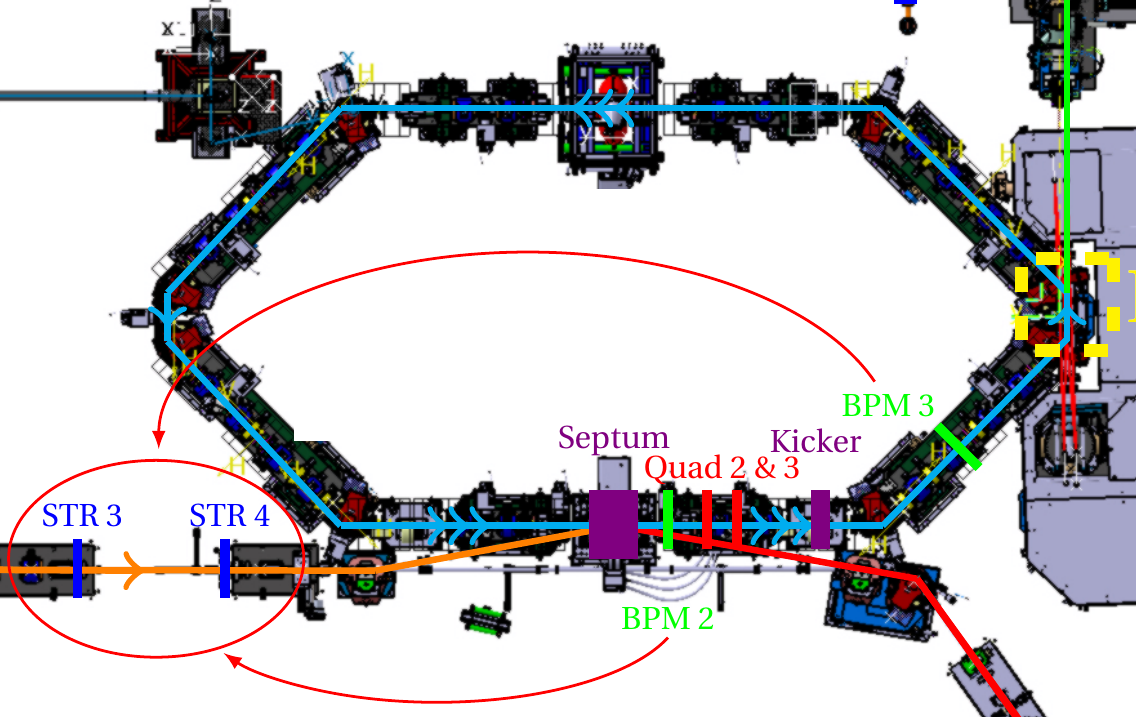}
   \caption{Overview of the ThomX's ring with localisation of the elements used in the injection feedback code.}
   \label{fig:injection_feedback}
\end{figure}

On ThomX the measured parameters are the beam positions in the two first Beam Position Monitors (BPMs) - named BPM2 and BPM3 - and the corrections are the kicker's kick value and the deviation of the two last steerers of the TL - named STR3 and STR4 - as shown in Fig.~\ref{fig:injection_feedback}.

To compute the steerers impact on the beam positions at BPMs' localisation it was chosen to calculate the equation of propagation using the classical linear transfer matrix calculation with parametrization of steerers' and kicker's values.

From BPMs' x and y values, kicker's kick value and the equation of propagation from BPM2 to BPM3, one may compute the vector (x,px,y,py) representing the transverse parameters of the beam's centroid at the localisation of the BPM2. 

The same may be done with the BPMs' wanted values - resumed in Table~\ref{tab:wanted_val} - to correctly inject the beam.

\begin{table}[!hbt]
   \centering
   \caption{Values Wanted in BPM2 and BPM3}
   \begin{tabular}{lcc}
       \toprule
                       & \textbf{BPM2}   & \textbf{BPM3} \\
       \midrule
           x & \SI{-8.9}{\milli\meter}  &  \SI{0}{\meter}                      \\ %[3pt]
           y & \SI{0}{\meter} & \SI{0}{\meter}                   \\ 
           px & & \SI{0}{\radian} \\ \bottomrule
   \end{tabular}
   \label{tab:wanted_val}
\end{table}

In BPM3, px is imposed as 0 to smooth the injection and to calculate the  kick value wanted for the kicker.

Finally, the steerers' deviation wanted are computed such that the inverse propagation from the BPM2 to the STR3 using wanted transverse beam's centroid parameters is equal to the same propagation but with measured beam's parameters and used steerers' deviations.

To test those computations, some simulations have been done.

\section{Injection Feedback Test}

The test of the feedback uses simulations done with MadX.
A particle - represented by its transverse parameters x,px,y and py - is selected randomly within the transverse ellipse at 5~$\sigma$ of the beam at the beginning of the TL (see paper~\cite{loss_map}) as predicted by linac simulations. 
This particle is tracked along TL and first ring turn to simulate the propagation of the beam's centroid.
An evaluation of x and y values at the BPM2 and BPM3 allows to compute the first wanted corrections.
To avoid overcorrections only some percentage of the differences between used deviations and computed ones are applied for the next iteration - see Table~\ref{tab:percent} - and the process is repeated with the same beam's centroid. 

\begin{table}[!hbt]
   \centering
   \caption{Percentage of Correction Applied. $\delta C$ Is the Maximum of Differences between Each Used and Computed Deviations}
   \begin{tabular}{ccc}
       \toprule
            \textbf{$\delta C$ > \SI{e-4}{}}   & \textbf{$\delta C$  > \SI{e-5}{}} & \textbf{$\delta C$ < \SI{e-5}{}} \\
       \midrule
           10\% & 20\% & 100\% \\ \bottomrule
   \end{tabular}
   \label{tab:percent}
\end{table}

\subsection{Estimation of the Injection}
%\subsubsection{Estimator definition}

To estimate the quality of the injection, the estimator of the Eq.~\eqref{eq:estimator} is used.

\begin{equation}\label{eq:estimator}
	E_v =  \sqrt{\frac{\sum_{Ring's~BPMs}{v^2}}{\#~of~BPM}},
\end{equation}
where $v$ can be x, px, y or py and ring's BPMs are all BPMs of the ring first turn except the first one, excluded because of the off-axes travelling during the injection.

%\subsubsection{Convergence Criterion}
We consider that the injection simulation is good enough - hence that the injection's tests converge to a solution - if $E_x$ and $E_y$ are below \SI{10}{\micro\meter} and $E_{px}$ and $E_{py}$ are below \SI{10}{\micro\radian}. Those limits are chosen pretty small as this test does not yet take into account the beam's centroid fluctuations or the uncertainty on BPMs' measurements.

\subsection{Tests Results}

\begin{figure*}[!bth]
    \centering
    \includegraphics*[width=0.83\textwidth]{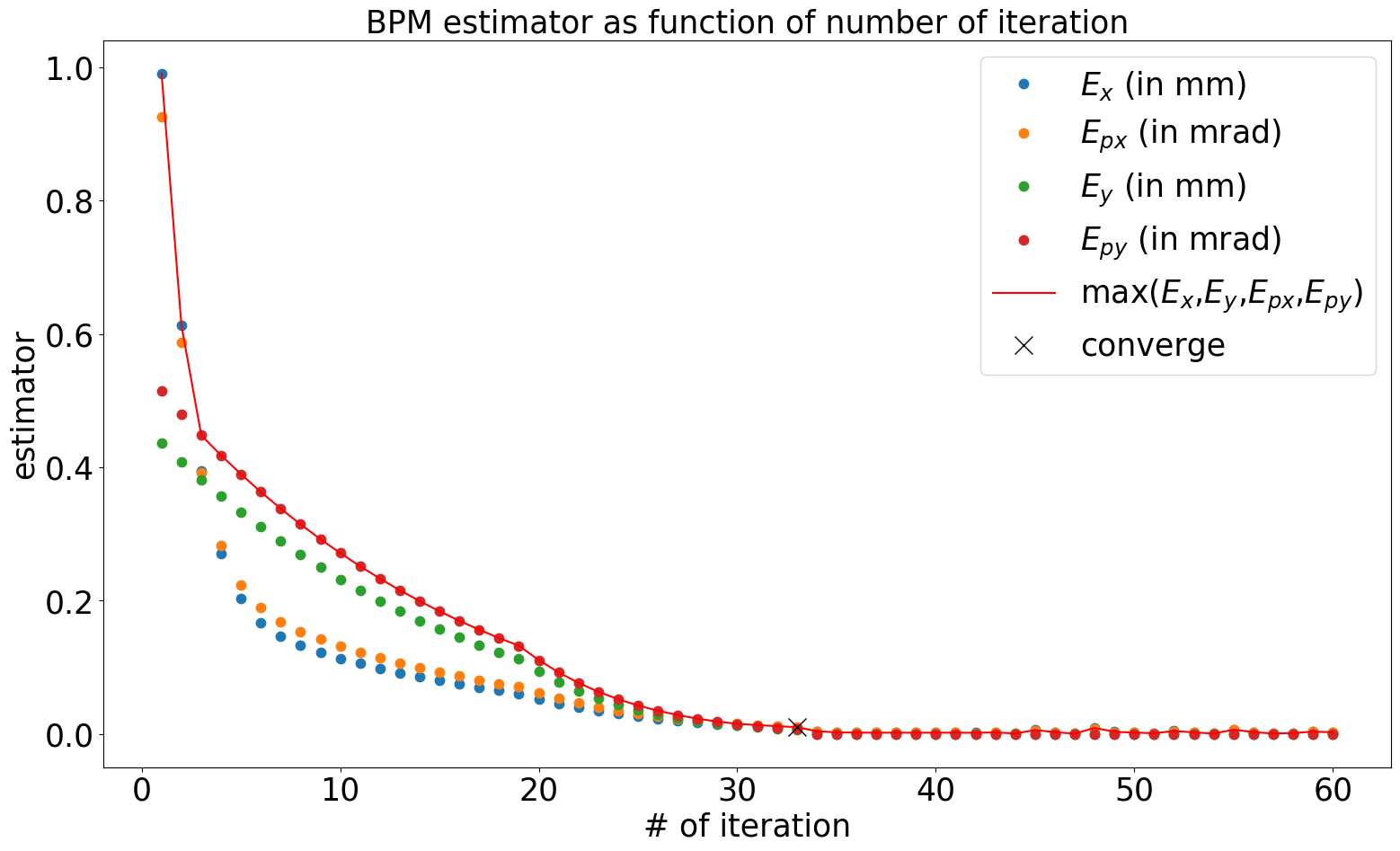}
    \caption{Evolution of the estimator at each iteration of the injection feedback code. The percentage of correction applied at each step depends on the maximum differences between each used and computed deviations $\delta C$ as follows: $\delta~C$~>~\SI{e-4}{}~$\rightarrow$~10\%,~  $\delta~C$~>~\SI{e-5}{}~$\rightarrow$~20\%,~ $\delta~C$~<~\SI{e-5}{}~$\rightarrow$~100\%.
    }
    \label{fig:one_optimisation}
\end{figure*}

Figure~\ref{fig:one_optimisation} shows the evolution of the estimator over 60 iterations.
33 iterations - hence less than a minute on ThomX - have been needed to reach the convergence criteria and it stays verified during the 27 next iterations.

%\begin{figure}[!htb]
%   \centering
%   \includegraphics*[width=0.75\columnwidth]{plot/BPM_estimator_as_function_of_number_of_iteration_54.png}
%   \caption{Evolution of the estimator at each iteration of the injection feedback code.}
%   \label{fig:one_optimisation}
%\end{figure}

\begin{figure*}[!bth]
    \centering
    \includegraphics*[width=0.98\textwidth]{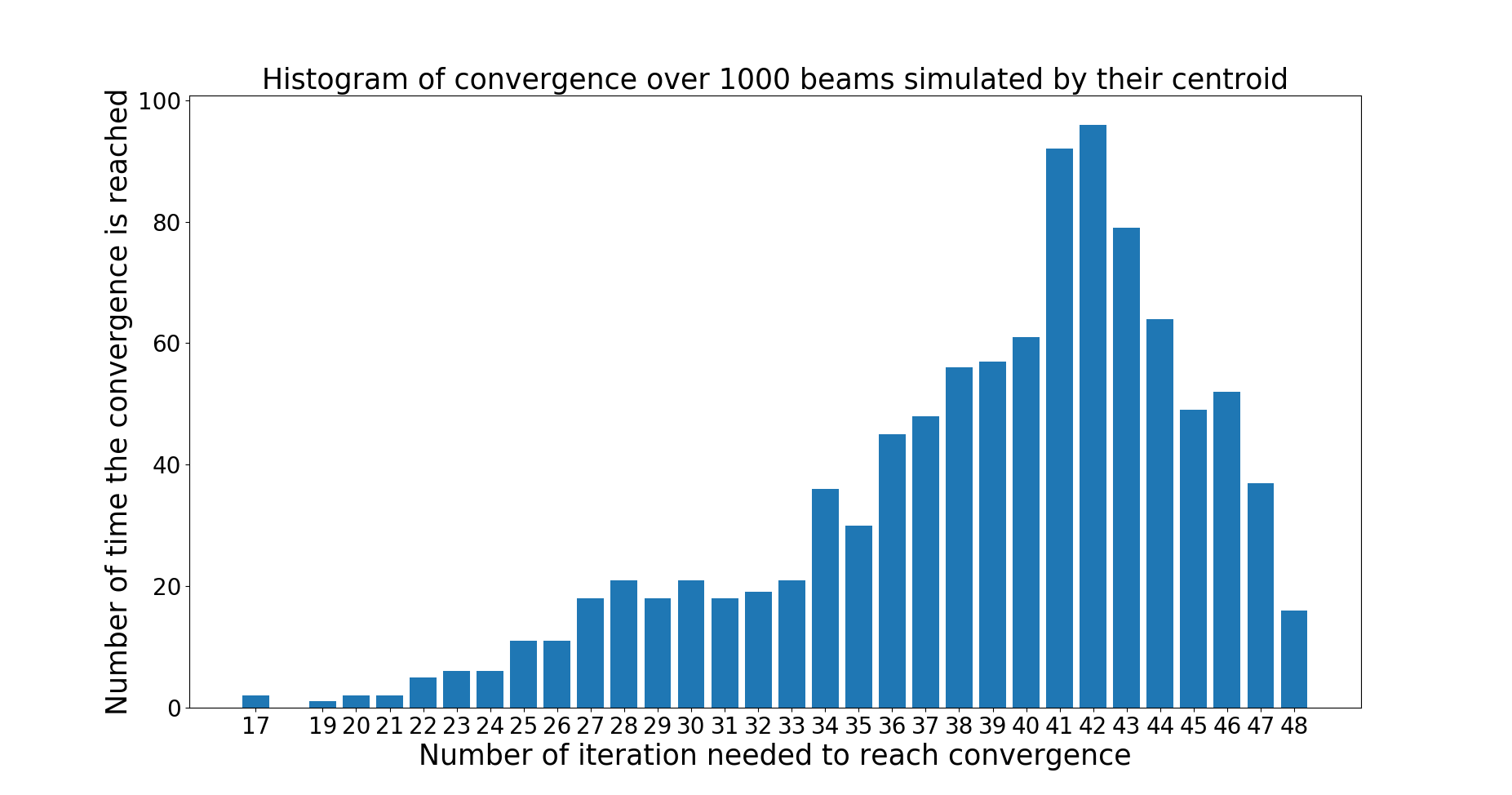}
    \caption{Histogram of the number of iteration needed to achieve the convergence's criteria. 1000 beams' centroid injections have been simulated. The percentage of correction applied at each step depends on the maximum differences between each used and computed deviations $\delta C$ as follows: $\delta~C$~>~\SI{e-4}{}~$\rightarrow$~10\%,~  $\delta~C$~>~\SI{e-5}{}~$\rightarrow$~20\%,~ $\delta~C$~<~\SI{e-5}{}~$\rightarrow$~100\%.}
    \label{fig:histo_iteration}
\end{figure*}

This test of feedback code has been done over 1000~times and a histogram of number of iteration to reach convergence is presented in Fig.~\ref{fig:histo_iteration}. During those tests the criteria of convergence is always reached in less than 48 iterations.

Moreover, MadX allows to recover losses of particles during tracking and no losses have been encountered during feedback operations.

\section{Code Robustness}

Some simulations have been done with small variations of the definition of ThomX's line on MadX - element's displacement and misalignment - but without modification of the propagation equations.
For those simulations the feedback system converges anyway but the number of iterations may increase up to a 100, which means a few minutes on ThomX.

This code's robustness justifies the use of linear propagation equations while the physical beam's propagation is not truly linear.

\section{CONCLUSION}

A feedback system has been developed for the injection in the ThomX ring.
Some preliminary tests have shown good and robust results. Further investigations have to be done with beam's fluctuations and BPMs' uncertainty to validate this behaviour.
Once ThomX commissioning starts this feedback will be applied to the real machine.

\section{ACKNOWLEDGEMENTS}

THOMX is financed by the French National Research Agency under the EQuipex program ANR-EQPX-51.
%Any acknowledgement should be in a separate section directly preceding
%the \textbf{REFERENCES} or \textbf{APPENDIX} section.

%\section{APPENDIX}
%Any appendix should be in a separate section directly preceding
%the \textbf{REFERENCES} section. If there is no \textbf{REFERENCES} section,
%this should be the last section of the paper.

%
% only for "biblatex"
%
\ifboolexpr{bool{jacowbiblatex}}%
	{\printbibliography}%
	{%
	% "biblatex" is not used, go the "manual" way
	
	%\begin{thebibliography}{99}   % Use for  10-99  references
	
} % end \ifboolexpr
%
% for use as JACoW template the inclusion of the ANNEX parts have been commented out
% to generate the complete documentation please remove the "%" of the next two commands
% 
%%%\newpage

%%%\include{annexes-A4}

\end{document}